\documentclass[aps,prl,floatfix,twocolumn,superscriptaddress,nofootinbib,a4paper]{revtex4-1}
\pdfoutput=1

\usepackage{amsmath}
\usepackage{amssymb}
\usepackage{graphicx}
\usepackage{hyperref}
\usepackage{color}
\usepackage{xcolor}
\usepackage[T1]{fontenc} 
\usepackage{slashed}

\usepackage[normalem]{ulem} 

\definecolor{blus}{cmyk}{1,1,0,0.6}
\hypersetup{colorlinks,bookmarksopen,bookmarksnumbered,linkcolor=blus,pdfstartview=FitH,urlcolor=blus,citecolor=blus}

\allowdisplaybreaks

\newcommand{\AddrOXF}{%
Rudolf Peierls Centre for Theoretical Physics, University of Oxford, Parks Road, Oxford OX1 3PU, UK
}

\begin{document}

\title{Electroweak mass difference of mesons}

\author{Antonio Pich}\email{pich@ific.uv.es}
\affiliation{Departament de F\'isica Te\`orica, IFIC, Universitat de Val\`encia – CSIC,
Parque Cient\'ifico, Catedr\'atico Jos\'e Beltr\'an 2, E-46980 Paterna, Spain}
\author{Arthur Platschorre}\email{arthur.platschorre@physics.ox.ac.uk}
\affiliation{\AddrOXF}

\author{Mario Reig}\email{mario.reiglopez@physics.ox.ac.uk}
\affiliation{\AddrOXF}

\begin{abstract}\noindent
We consider electroweak (EW) gauge boson corrections to the masses of pseudoscalar mesons to next to leading order (NLO) in $\alpha_s$ and $1/N_C$. The pion mass shift induced by the $Z$-boson  is shown to be  $m_{\pi^\pm}-m_{\pi^0} = -0.00201(12)
$ MeV. While being small compared to the electromagnetic mass shift, the prediction lies about a factor of $\sim 4$ above the precision of the current experimental measurement, and a factor $O(10)$ below the precision of current lattice calculations.  This motivates future implementations of these EW gauge boson effects on the lattice. 
Finally, we consider BSM contributions to the pion mass difference.

\end{abstract}

\maketitle

\section{Introduction}
At very low energies, the strong interaction of mesons is successfully described by the chiral Lagrangian, a perturbative expansion in derivatives of the Goldstone fields and light quark masses. 
The effective action is entirely determined by the symmetries, and once the parameters of the theory are fixed by observation of several meson quantities, a highly predictive theory emerges, chiral perturbation theory \cite{Gasser:1984gg,Ecker:1994gg,Pich:1995bw}.

 In QCD with 3 light flavours, the global symmetry is $SU(3)_L\times SU(3)_R$, giving 8 Goldstone bosons after spontaneous symmetry breaking by the formation of quark condensates. Turning on quark masses, 
 $M_q=\text{diag}(m_u,m_d,m_s)$, explicitly breaks the flavour symmetry and the meson fields get a mass. The effective action does not allow one to obtain the meson masses purely as a function of quark masses, but it is possible to find relations that connect ratios of the meson masses to (renormalization-scheme independent) ratios of quark masses, one example being the renowned Gell-Mann-Oakes-Renner relation $\frac{m^2_{K^\pm}-m^2_{K^0}}{m^2_\pi}=\frac{m_u-m_d}{m_u+m_d}$.

 The process of gauging part of the global symmetries also breaks the chiral flavour symmetry, generating masses for the pseudoscalar mesons. This is well-known for the case of electromagnetism (EM) which breaks the shift symmetries of the charged mesons, thereby generating the pion and kaon mass shifts: $\delta m_\pi= m_{\pi^\pm}-m_{\pi^0}$. 
This quantity has been computed using current algebra \cite{Das:1967it} and in chiral perturbation theory with explicit resonance fields \cite{Ecker:1988te}, giving $\delta m_\pi$ compatible with the experimental result \cite{Crawford:1990jc}, 
\begin{equation}\label{eq:exp_mass_shift}
    \delta m_\pi|_{\mathrm{exp}} = m_{\pi^\pm} - m_{\pi^0} = 4.5936\pm0.0005 \text{ MeV}\,.
\end{equation}
%

 The pion mass shift is a quantity that can also be computed on the lattice. This direction was initiated in \cite{PhysRevD.76.114508} and currently has reached a level of considerable accuracy \cite{PhysRevLett.128.052003,Frezzotti:2022dwn}. The 
 most precise lattice result  \cite{PhysRevLett.128.052003}: 
\begin{equation}\label{eq:lattice_result}
    \delta m_\pi = m_{\pi^\pm} - m_\pi^0= 4.534\pm 0.042\pm 0.043
    \text{ MeV}\,,
\end{equation}
is compatible with the experimental measurement in Eq.~\ref{eq:exp_mass_shift}. While the error on the lattice still has to be substantially reduced to reach the experimental precision, given the rate of improvement of lattice precision in recent years it is not unreasonable to think that in a near future the size of both errors might be comparable. 

In this letter we show that heavy EW gauge bosons induce small, but possibly \textit{observable} mass shifts between the neutral and charged mesons, for both the pion and the kaon. 
Due to the chiral structure of the weak interaction, to leading order (LO) in $G_F$, only the $Z$ boson contributes to the mass shifts. Similar results to LO in $\alpha_{s}$ were noted in \cite{Knecht:1998sp}.

 By doing a calculation at NLO in both $\alpha_{s}$ and $1/N_{c}$, our results will show that the expected mass shift induced by the $Z$ lies well above the uncertainty of the current experimental measurement and slightly below the lattice uncertainties.
This implies that future lattice simulations
should be sensitive to the effects of the EW gauge bosons, reflecting the need for an implementation on the lattice. This direction is particularly interesting to learn about flavour symmetry breaking by the weak interaction in the chiral limit. 
Finally, we discuss future directions including effects of new physics on the mass differences of mesons.

\section{Electroweak interaction and the pion mass difference}
QCD with 3 light flavours has a $SU(3)_L\times SU(3)_R$ global flavour symmetry.  Starting at order $O(p^2)$, and neglecting momentarily  quark masses, the effective Lagrangian below the chiral symmetry breaking scale is of the form:
\begin{equation}\label{eq:lag-p2}
    \mathcal{L}_2 = \frac{F^{2}}{4
    }\mathrm{Tr} \left( D^\mu U \left(D_\mu U\right)^{\dagger}\right)\,,
\end{equation}
where $F$ is the chiral coupling constant and 
the $SU(3)$ matrix $U = \mathrm{exp}\left[i \frac{\sqrt{2}}{F} \Phi \right]$
incorporates the pseudoscalar Goldstone octet
\begin{equation}
\Phi = \begin{pmatrix}
    \frac{\pi^{0}}{\sqrt{2}} + \frac{\eta^{0}}{\sqrt{6}} & \pi^{+} & K^{+} \\ \pi^{-} & -\frac{\pi^{0}}{\sqrt{2}} + \frac{\eta^{0}}{\sqrt{6}} & K^{0} \\ K^{-} & \overline{K}^{0} & - \frac{2}{\sqrt{6}} \eta^{0} 
\end{pmatrix} \,.
\end{equation}
%

%

In the SM, the $SU(2)\times U(1)$ subgroup of this flavour symmetry is gauged. In general, gauging a subgroup of $SU(3)_{L}\times SU(3)_{R}$ by gauge bosons $L$ and $R$ is done by introducing a covariant derivative of the form:
\begin{equation}
    D_{\mu} U = \partial_{\mu} U - i Q_{L}  \ell_{\mu}U + i   U r_{\mu}Q_{R}\,.
\end{equation}
For the SM gauge bosons this amounts to introducing:
\begin{widetext}
\begin{align}\label{eq:cov_der}
    D_\mu U =& \partial_{\mu} U - i \frac{g}{\sqrt{2}} \left(W^{+}_{\mu} T^{-}_{W} +  W^{-}_{\mu} T^{+}_{W}\right) U   - ie \left( A_{\mu} - \tan{\theta_{W}} Z_{\mu} \right)[Q_{\mathrm{em}},U]  - i \frac{g}{\cos{\theta_{W}}} Z_{\mu} T_{3L}  U \,,
\end{align}
\end{widetext}
where we have explicitly included the photon and the EW gauge bosons with the generators:
\begin{equation}
 T_{W}^{-} = \left(T^{+}_{W}\right)^{\dagger}= \begin{pmatrix}
   0 & V_{ud} & V_{us}  \\ 0 & 0 & 0 \\  0 & 0 & 0 \\ \end{pmatrix}
\,, 
\end{equation}
and the diagonal matrices $T_{3L}=\text{diag}(1/2,-1/2,-1/2)$ and $Q_{\mathrm{em}}=\text{diag}(2/3,-1/3,-1/3)$. The heavy EW gauge bosons are introduced as spurions in order to track the pattern of explicit symmetry breaking. However, since these particles lie well above the cut-off of the effective theory, usually taken to be $\Lambda_{\chi\mathrm{SB}}\sim4\pi F$, special care has to be taken in deriving explicit results from this Lagrangian. We shall return to this issue momentarily. 

Expanding Eq.~\ref{eq:lag-p2} to quadratic order in $\Phi$, we can see that non-zero Goldstone masses are
generated by terms of the form:
\begin{equation}
- \frac{F^{2}}{2} \mathrm{Tr}\left( Q_L U Q_R U^\dagger \right) \dot= \frac{1}{2}  
    \mathrm{Tr}\left([Q_L,\Phi][\Phi,Q_R] \right) 
\end{equation}
where $Q_L$ and $Q_R$ are  spurion matrices representing the action of gauge fields.

One notices that not all of these terms are breaking the shift symmetries in the chiral limit, because meson self-energies are generated by loop diagrams with no external gauge bosons.
Consequently, terms involving different gauge bosons do not contribute at LO to the meson masses. Since the $W^\pm$ couplings are purely left-handed, they cannot contribute to $Q_R$ and, therefore, do not generate any meson mass shift.
%

The only contribution to $Q_R$ comes from the spurion $Q_{\mathrm{em}}$, which as seen from Eq.~\ref{eq:cov_der} occurs for both the photon and the $Z$, and acts as:
\begin{equation}
  [Q_{\mathrm{em}},\Phi] =  
    \begin{pmatrix}
 0 & \pi^{+} & K^{+} \\ -\pi^{-} & 0 & 0 \\ -K^{-} & 0 & 0
\end{pmatrix}\,.
\end{equation}
This implies that only charged mesons can get a mass and this occurs through the interaction with neutral gauge bosons, which contribute as:
\begin{equation}\label{eq:contrib_1}
\frac{eg }{2\cos{\theta_{W}}} \mathrm{Tr}\left(  \left[T_{3L}
   , \Phi \right] [\Phi,Q_{\mathrm{em}}]\right)(A_{\mu} - \tan{\theta_{W}} Z_{\mu})Z^{\mu}\,,
\end{equation}
and:
\begin{equation}\label{eq:contrib_2}
    \frac{e^{2}}{2} \mathrm{Tr}\left([Q_{\mathrm{em}},\Phi][\Phi,Q_{\mathrm{em}}] \right)(A_{\mu} - \tan{\theta_{W}} Z_{\mu})(A^{\mu} - \tan{\theta_{W}} Z^{\mu})\,.
\end{equation}
Again, the term involving $A_\mu Z^\mu$ cannot contribute to meson masses. 
Combining Eq.~\ref{eq:contrib_1} and Eq.~\ref{eq:contrib_2}, and retaining only the relevant terms involving $A_\mu A^\mu$ and $Z_\mu Z^\mu$, the interaction 
reads:
\begin{equation}\label{eq:charged_meson-gauge_boson_interaction}
    e^{2}\left(  \pi^{+} \pi^{-} + K^{+} K^{-} \right) (A_{\mu}A^\mu - Z_{\mu}Z^\mu)\,.
\end{equation}
An order of magnitude estimate can be given at this point for the $Z$-boson induced mass shift using naive dimensional analysis:
\begin{equation}
    \Delta m_\pi^2=\frac{e^2}{4 \pi^{2} M_Z^2}\Lambda_{\chi\mathrm{SB}}^4\rightarrow \delta m_\pi\sim  0.002
    \text{ MeV}\,.
\end{equation}
The fact that this estimate lies above the current experimental uncertainty and is comparable to the lattice precision motivates us to perform a more careful analysis.

As 
in
the electromagnetic (EM) contribution \cite{Ecker:1988te},  we capture the effects of both $A_\mu$ and $Z_\mu$ by adding the following local operators involving the spurion matrices
$Q_{\mathrm{em}}$ and $Q_{L,R}^Z\equiv \frac{g}{\cos{\theta_W}} \mathcal{Q}_{L,R}$:
%
\begin{equation}
\mathcal{L}_2^C=e^2C_{em}\langle Q_{\mathrm{em}}UQ_{\mathrm{em}}U^\dagger\rangle + 4 \sqrt{2}G_{F}C_{Z}
\langle \mathcal{Q}_{L} U \mathcal{Q}_{R} U^\dagger\rangle\, ,
\label{lowenergyconstants}
\end{equation}
with $4\sqrt{2}G_F$ the low-energy coupling of the $Z$ boson, 
\begin{align}
& \mathcal{Q}_{L} =\begin{pmatrix}
  \frac{1}{2} - \frac{2}{3}x & 0 & 0 \\ 0 & -\frac{1}{2} + \frac{1}{3}x & 0  \\ 0 & 0 & - \frac{1}{2} + \frac{1}{3} x \end{pmatrix}\,,\,\, \qquad \\& \mathcal{Q}_{R} =  \begin{pmatrix}
  - \frac{2}{3}x & 0 & 0 \\ 0 & \frac{1}{3}x & 0  \\ 0 & 0 & \frac{1}{3}x\end{pmatrix} 
\end{align}
and $x = \sin^{2}\theta_{W}$.
The determination of $C_Z$ to NLO in $\alpha_{s}$ and $1/N_{c}$ is the goal of this letter. 
 
The coefficients $C_{\mathrm{em}}$ and $C_Z$ are low-energy constants 
determined from the high-energy theory
%
and determine the 
electromagnetic and electroweak meson mass differences $\Delta m^2_P \equiv m^2_{P^\pm}-m^2_{P^0}$ of pions and kaons in the chiral limit:
\begin{equation}\label{eq:pion_mass_sq}
    \Delta m^2_\pi = \Delta m^2_K = \frac{2 e^2}{F^2}\left( C_{\mathrm{em}} - \frac{C_Z}{M_Z^2}\right) .
\end{equation}

In \cite{Ecker:1988te} it was shown
that the EM mass shift from resonance exchange saturates the constant $C_{\mathrm{em}}$ and is given in terms of the resonance parameters $F_V, M_V$  by:
\begin{equation}
     \Delta m^2_\pi|_{\mathrm{em}}=\frac{3\alpha_{\mathrm{em}}}{4\pi F^2}F_V^2M_V^2\ln \frac{F_V^2}{F_V^2-F^2}\,.
\end{equation}
A corresponding resonance loop calculation including the $Z$ boson in order to determine $C_Z$ is subtle.
The reason is that the parameter $M_Z$ lies well above the cut-off, 
$\Lambda_{\chi\mathrm{SB}}$, and the $Z$ therefore must be integrated out. 

The resulting EFT is QCD with four-fermion operators that encode all the information of the chiral symmetry breaking by the EW bosons. 
Using the renormalization group (RG) to run 
the Wilson coefficients of these operators down to a scale $\mu \sim 1 \ \text{GeV}$ allows matching to the operators in Eq.~\ref{lowenergyconstants} 
of the chiral Lagrangian and thereby a determination of $C_{Z}$. 
\subsection{Z-induced left-right four quark operators}
%
Integrating out the $Z$ boson 
introduces 4-fermion operators that break the chiral $SU(3)_L\times SU(3)_R$ symmetry.  The relevant 
left-right (LR) operators are:
\begin{align}\label{eq:4-fermion_ops}
   &[Q^{LR}_{1}]_{ijk\ell} =  \left(\overline{q}_{Li} \gamma^{\mu} q_{Lj} \right) \left( \overline{q}_{Rk} \gamma^{\mu} q_{R\ell} \right)\qquad \\&
   [Q_{2}^{LR}]_{ijk\ell} = \left(\overline{q}_{Li} q_{Rk} \right)\left(\overline{q}_{R\ell} q_{Lj} \right)\,,
\end{align}
with $i,j,k,\ell$ being light-quark flavour indices. While $Q_{1}^{LR}$ is generated by a $Z$-exchange at tree level, $Q^{LR}_{2}$ is obtained after applying a Fierz-identity on the gluon corrections to $Q_{1}^{LR}$.

The effective lagrangian below $M_Z$ reads: 
\begin{equation}
    \mathcal{L}_{\mathrm{eff}} = - 4 \sqrt{2} G_{F} \sum_{ijk\ell} \left(\mathcal{Q}_{L}\right)_{ij} \left(\mathcal{Q}_{R}\right)_{k\ell} \left[ C_{1} Q_{1}^{LR} + C_{2} Q_{2}^{LR} \right]_{ijk\ell}\,,
    \label{introoperators}
\end{equation}
with $C_{1,2}$ being the Wilson coefficients. 

When QCD effects
are taken into account, the renormalised Wilson coefficients at the $M_Z$ scale become \cite{Buras:2012fs}:
\begin{align}
&C_{1} = 1 + \frac{\alpha_{s}}{4\pi} \frac{3}{N_{c}} \left[- \ln{\frac{M_{Z}^{2}}{\mu^{2}}} - \frac{1}{6} \right] \, ,\qquad 
\\&
C_{2} = \frac{\alpha_{s}}{4\pi} \left[- 6\ln{\frac{M_{Z}^{2}}{\mu^{2}}} - 1 \right]\,,
 \end{align}
where the non-logarithmic corrections are scheme dependent. The operators above will mix under RG flow and their evolution down to the scale of interest ($ \sim 1 \ \mathrm{GeV}$) can be calculated by standard procedures \cite{Buchalla:1995vs}, using their anomalous dimension matrices:
\begin{equation}\label{eq:differentialeq}
    \frac{d\Vec{C}}{d\ln{\mu}} = \gamma^{T} \Vec{C}\,.
\end{equation}
Up to order $O(\alpha_{s}^{2})$, this matrix can be expanded as:
\begin{equation}
    \gamma = \frac{\alpha_{s}}{4\pi} \gamma^{0} + \left(\frac{\alpha_{s}}{4\pi}\right)^{2} \gamma^{1} + O(\alpha_s^3)\,,
\end{equation}
with $\gamma^0,\gamma^1$ given by
\cite{Buras:2000if}:
\begin{widetext}
\begin{align}
    \gamma^{0} &= \begin{pmatrix} \frac{6}{N_{c}} & 12 \\ 0 & -6N_{c} + \frac{6}{N_{c}} \end{pmatrix} \,,\,\,\,\,\,\,
    \gamma^{1} = \begin{pmatrix} \frac{137}{6} + \frac{15}{2N^{2}_{c}} - \frac{22}{3N_{c}}f & \frac{200}{3}N_{c} - \frac{6}{N_{c}} - \frac{44}{3}f \\ \frac{71}{4}N_{c} + \frac{9}{N} - 2f & - \frac{203}{6} N_{c}^{2} + \frac{479}{6} + \frac{15}{2N^{2}_{c}} + \frac{10}{3} N_{c} f - \frac{22}{3N_{c}}f \end{pmatrix} \,.
\end{align}
\end{widetext}
Solving 
Eq.~\ref{eq:differentialeq} yields the evolution:
\begin{equation}
    \Vec{C}(\mu) = T \;\mathrm{exp}\left[ \int_{\alpha_{s}(M_{Z})}^{\alpha_{s}(\mu)} d\alpha_{s} \frac{\gamma^{T}}{\beta(\alpha_{s})}\right] \Vec{C}(M_{Z})\,,
\end{equation}
where we have introduced the QCD $\beta$ function as:
\begin{equation}
    \beta = -2 \alpha_{s} \left[\beta_{0} \frac{\alpha_{s}}{4\pi} + \beta_{1} \left(  \frac{\alpha_{s}}{4\pi}\right)^{2}+ O(\alpha_{s}^{3}) \right]\,.
\end{equation}
The coefficients used are given by $\beta_{0} = \frac{11N_{c} - 2f}{3}$ and $\beta_{1} =  \frac{34}{3} N_{c}^{2} - \frac{10}{3}N_{c}f - \frac{N_{c}^{2}-1}{N_{c}}f$ \cite{ParticleDataGroup:2022pth} where $f$ is the number of active flavours. 

To NLO
and 
after integrating out the $b$ and $c$ quark, the Wilson coefficients at the scale $\mu \sim 1 \ \mathrm{GeV}$ are: 
\begin{align}\label{eq:wilson_1GeV}
    C_{1} &= 0.92\,,\,\,\,\,\, C_{2} = -2.45 \, .
\end{align}
Similar enhancements of $C_{2}$ are noticed in \cite{Buras:2001ra}.

\subsection{\boldmath Matching to the chiral Lagrangian at large $N_c$}
%
We proceed to match the resulting EFT to the chiral Lagrangian. We do so by calculating the expectation value of the matrix elements of the 4-fermion operators in the large-$N_c$ limit in which products of colour-singlet currents factorise. 

In this limit, the operator $Q^{LR}_{1}$ reduces to the product of a left and a right currents:
%
\begin{equation}\label{eq:Q1LR}
[Q^{LR}_{1}]_{ijk\ell} =  \mathcal{J}^{\mu}_{L,ji}\;\mathcal{J}^R_{\mu ,\ell k} \, .
\end{equation}
Since the low-energy representation of these currents starts at $O(p)$ in the chiral-perturbation-theory expansion, the large-$N_C$ expression of $Q^{LR}_{1}$ is of $O(p^2)$ and, therefore, does not contribute to the $O(p^0)$ operator in Eq.~\ref{lowenergyconstants}.
Owing to its different scalar-pseudoscalar structure, the operator $Q^{LR}_{2}$ does contribute at
$O(p^0)$, receiving a chiral enhancement of the form:
\begin{align}\label{eq:q2}
   [Q^{LR}_{2}]_{ijk\ell} &=  \langle \overline{q}^{i}_{L} q^{k}_{R}   \rangle \langle \overline{q}^{\ell}_{R} q^{j}_{L} \rangle\, \left\{1 + O\left(\frac{1}{N_{c}}\right) \right\}\\& = \frac{1}{4}B^{2}_{0} F^{4} U_{ki} U^{\dagger}_{j \ell}\, \left\{ 1 +  O\!\left(\frac{1}{N_{c}}\right)\right\} + O\!\left(p^2\right)\,,
   \label{q2}
\end{align}
with 
$B_0 = - \langle \bar{q}q\rangle /F^2 = m_{\pi^\pm}^2/(m_u+m_d)$.

Matching the contribution of $Q_2^{LR}$
to the effective theory, 
 a LO estimate in $N_c$ can be given for 
 $C_Z$:
\begin{equation}
    C_{Z} = -\frac{1}{4}\, B^{2}_{0}(\mu)\, F^{4} \, C_{2}(\mu) \,.
\label{eq.CZQ2}
\end{equation}
One can easily check that, in the large-$N_c$ limit, the $\mu$ dependence of $C_2(\mu)$ is exactly cancelled by the quark-mass factors in $B_0^2(\mu)$, as it should.



\subsection{\boldmath $1/N_c$ corrections to $Q_1^{LR}$}
%
As shown in \cite{Knecht:1998sp}, the low-energy constants in Eq. \ref{lowenergyconstants} can be related to the two-point correlation function of a left and a right QCD currents, $\Pi_{LR}(Q^2)$, which converges nicely in the UV. This fact allows one to evaluate the leading non-zero $O(p^0)$ contributions of $Q^{LR}_{1}$, originating from loops of Goldstone bosons and vector and axial-vector resonance fields, which are NLO corrections in $1/N_c$.
The full details of the calculation are given in the Appendix. Integrating only the low-energy region $0\le Q^2\le\mu^2$ (contributions from $Q^2 > \mu^2$ are already included in the Wilson coefficients), one finds 
\begin{widetext}
\begin{equation}
 \left. \Delta C_Z \right|_{Q^{LR}_{1}}  =
 \frac{3}{32\pi^2} \left\{
\sum_A F_{A_i}^2M_{A_i}^4\log \left ( 1+\frac{\mu^2}{M_{A_i}}\right ) - 
 \sum_V F_{V_i}^2M_{V_i}^4 \log \left ( 1+\frac{\mu^2}{M_{V_i}}\right ) \right\} C_1(\mu)\,.
\label{q1}  
\end{equation}            
\end{widetext}
Since we are interested in the matrix element of the operator $Q_1^{LR}$ at around the $\mu\sim 1$~GeV scale, we work in the lightest-resonance approximation with their couplings fixed through the Weinberg conditions \cite{Weinberg:1967kj,Pich:2002xy}:
\begin{equation}
    F_{V}^{2} = \frac{M_{A}^{2}}{M_{A}^{2} - M_{V}^{2}}\, F^{2}\, , \qquad  F_{A}^{2} = \frac{M_{V}^{2}}{M_{A}^{2} - M_{V}^{2}}\, F^{2}\, .
    \label{Weinbergsum}
\end{equation}
Within the single-resonance approximation that we have adopted, $M_{A} = \sqrt{2} M_{V}$ \cite{Pich:2002xy}. For the numerical evaluation we will take $M_{V} = M_{\rho} = 775.26 \pm 0.23$ MeV and $F = F_{\pi} = 92.1 \pm 0.8$ MeV \cite{ParticleDataGroup:2022pth}. 
As expected from its loop suppression, $\left.\Delta C_Z\right|_{Q^{LR}_{1}}$
is of $O(F^2)\sim O(N_c)$ and, therefore, is a NLO correction in $1/N_c$ of about $O(10\%)$ with respect to the leading $O(F^4)\sim O(N_c^2)$ contribution from $Q_2^{LR}$ in Eq.~\ref{eq.CZQ2}.


\subsection{EW contribution to the pion mass difference}
%
%
%

Using 
Eq.~\ref{eq:pion_mass_sq} and the results above in Eqs. \ref{eq.CZQ2}, \ref{q1} and \ref{Weinbergsum}, the pion mass shift induced by the $Z$ reads:
%
%

%
\begin{widetext}
    \begin{align}
        \Delta m_\pi^2|_Z=\frac{e^2}{M_Z^2}\left\{ \frac{F^2}{2}B_0^2(\mu)C_2(\mu) + \frac{3}{16\pi^2}C_1(\mu)\frac{M_A^2M_V^2}{M_A^2-M_V^2}\left [ M_V^2\log \left (1+\frac{\mu^2}{M_V^2}\right ) - M_A^2\log \left (1+\frac{\mu^2}{M_A^2}\right ) \right ]  \right\} .
    \end{align}
\end{widetext}
This translates into a $Z$-induced pion mass difference:
\begin{align}\label{eq:result}
    &\delta m_\pi|_{Z}\approx \frac{\Delta m_\pi^2|_Z}{2m_\pi}= -0.00201(7)(2)(10) \text{ MeV}\,,
\end{align}
where we have used $m_\pi=134.9768 \pm 0.0005$ MeV \cite{ParticleDataGroup:2022pth} and 
$(m_u+m_d)/2 = 3.381 \pm 0.040$~MeV
\cite{FlavourLatticeAveragingGroupFLAG:2021npn}. 
The first error displays the parametric uncertainty induced by the different inputs. The second uncertainty accounts for the renormalization-scale dependence in the interval
$\mu\in [0.8 , 1.2]$~GeV which, as shown in the figure, is tiny. We have added half the difference between the LO and NLO results as an estimate of unknown higher-order effects (third error).

\begin{figure*}[t]
	\centering
	\includegraphics[width=0.55\textwidth]{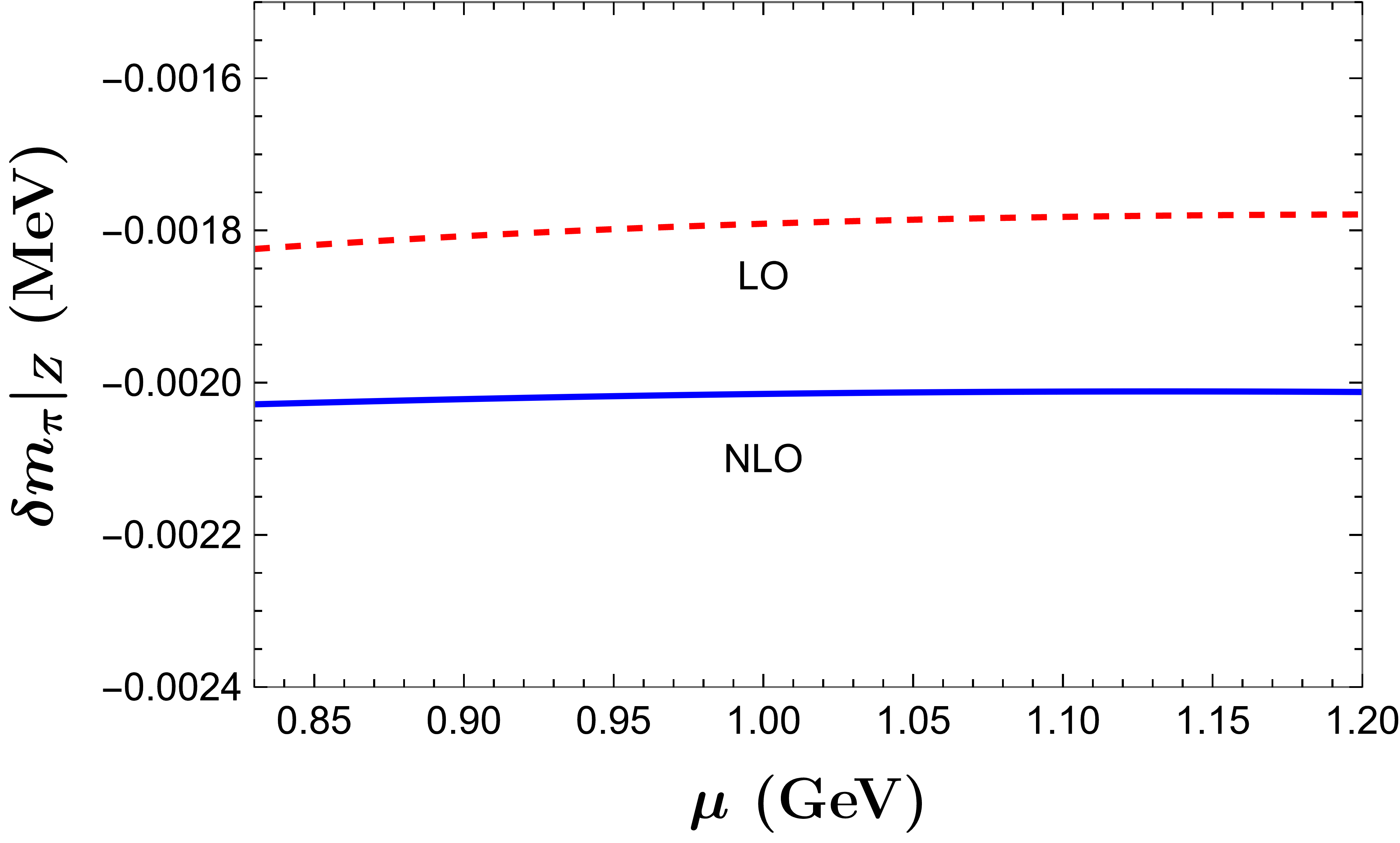}
		\caption{Renormalisation scale dependence of the pion mass shift induced by the $Z$ boson.}
	\label{fig:renormalisation_scale_dependence}
\end{figure*}

%
%
%
We notice that the $Z$-boson contribution is about a factor of $\sim 4$ larger than the experimental error in Eq.~\ref{eq:exp_mass_shift} and $\sim O(10)$ smaller than the current lattice precision in Eq.~\ref{eq:lattice_result}, reinforcing the motivation to incorporate these effects on the lattice. The renormalization scale dependence of this result for energies in the range $[0.8,1.2]$ GeV is plotted in  figure \ref{fig:renormalisation_scale_dependence}.  

\section{Discussion}
%

Before closing we comment on several points that deserve mention. 
\begin{itemize}
    \item The estimate in Eq.~\ref{eq:result} 
is based on a NLO evaluation of the Wilson coefficients $C_{1,2}(\mu)$, which
    depends on the precise values of the strong coupling at $M_Z$, $\alpha_{s}(M_Z) = 0.1184 \pm 0.0008$ \cite{FlavourLatticeAveragingGroupFLAG:2021npn}, and at the different matching scales (known to percent level or better).
    \item Our result $\delta m_\pi |_Z$ appears to be of the same order as the two-loop EM effect, which naively one expects to be: 
\begin{equation}
    \delta m_\pi|^{(2)}_{\mathrm{em}}\approx \left (\frac{\alpha_{\mathrm{em}}}{2\pi} \right )\delta m_\pi|^{(1)}_{\mathrm{em}}\,.
\end{equation}

\item BSM models that generate 4-quark LR operators at energies below the new physics scale, $\Lambda_{\mathrm{NP}}\gg \Lambda_{\chi\mathrm{SB}}$, will induce similar pion mass shifts. This is the case, for example, of the $Z^\prime$ models studied in \cite{Buras:2012fs}, and similar SM extensions. Since the QCD corrections dominate near the GeV scale, a reasonable estimate is just the rescaling:
\begin{equation}
    \delta m_\pi|_{\mathrm{NP}}=\frac{g_{\mathrm{NP}}^2}{\Lambda_{\mathrm{NP}}^2}\frac{\delta m_\pi|_Z}{4\sqrt{2} G_F}\,.
\end{equation}

If new physics is instead light, as proposed in \cite{DiLuzio:2021uty,Coyle:2023nmi}, one should rescale the resonance calculation for EM effects \cite{Ecker:1988te}.

\end{itemize}

\section*{Acknowledgments}
We would like to thank Prateek Agrawal, Hector Gisbert, Victor Miralles and Fernando Romero for helpful discussions and enlightening comments on the early drafts of this letter.
Antonio Pich is supported by Generalitat Valenciana, Grant No. Prometeo/2021/071, and
MCIN/AEI/10.13039/501100011033,             Grant No. PID2020-114473GB-I00.
Arthur Platschorre is supported by a STFC Studenship No. 2397217 and Prins Bernhard Cultuurfondsbeurs No. 40038041 made possible by the Pieter Beijer fonds and the Data-Piet fonds.

\section*{Appendix}

In the large-$N_C$ limit, the strong interaction reduces to tree-level hadronic diagrams. Keeping only those terms that are relevant for our calculation, the effective Lagrangian describing the mesonic world contains the LO Goldstone term $\mathcal{L}_2$ and the vector and axial-vector couplings (kinetic terms are omitted) \cite{Pich:2002xy}:
\begin{equation}
\mathcal{L}_{V,A}  = 
\sum_{V_i} \frac{F_{V_i}}{2\sqrt{2}}\,\langle V_i^{\mu\nu} f_{+\mu\nu}\rangle  + \sum_{A_i} \frac{F_{A_i}}{2\sqrt{2}}\,\langle A_i^{\mu\nu} f_{-\mu\nu}\rangle \, ,
\end{equation}
where $f_\pm^{\mu\nu} = u^\dagger F_L^{\mu\nu} u \pm u F_R^{\mu\nu} u^\dagger$ with
$U=u^2$ the Goldstone $SU(3)$ matrix and $F_{L,R}^{\mu\nu}$ the left ($\ell^\mu$)
and right ($r^\mu$) field strengths. The spin-1 resonances are described through the antisymmetric tensors $V_i^{\mu\nu}$ and $A_i^{\mu\nu}$ \cite{Ecker:1988te,Ecker:1989yg}.

The left and right QCD currents are easily computed, taking derivatives with respect to the external $\ell^\mu$ and $r^\mu$ fields:
\begin{eqnarray}\label{eq:Rcurrents}
\mathcal{J}^{\mu}_L &\!\! =  &\!\! 
i\frac{F^2}{2}\, D^\mu U U^\dagger + 
\sum_{V_i} \frac{F_{V_i}}{\sqrt{2}}\,\partial_\nu (u V_i^{\mu\nu} u^\dagger)
\nonumber\\ &&\!\!
+ \sum_{A_i} \frac{F_{A_i}}{\sqrt{2}}\,\partial_\nu (u A_i^{\mu\nu} u^\dagger)+\cdots
\end{eqnarray}
while $\mathcal{J}^{\mu}_R$ is obtained from this expression exchanging $u\leftrightarrow u^\dagger$ and putting a negative sign in the axial contributions.

The bosonization of $[Q_1^{LR}]_{ijk\ell}$ is formally given by \cite{Pich:1990mw}
\begin{equation}\label{eq:bosonization}
\langle [Q_{1}^{LR}(x)]_{ijkl}\rangle_G =
 \frac{\partial \Gamma}{\partial \ell_\mu^{ij}(x)}\, \frac{\partial \Gamma}{\partial r^{\mu,kl}(x)}
-i\,  \frac{\partial^2 \Gamma}{\partial \ell_\mu^{ij}(x)\,\partial r^{\mu,kl}(x)}
\end{equation}
with $\Gamma[\ell,r]$ the effective theory generating functional. The first term is just the product of the two currents and receives $O(p^0)$ contributions from loop diagrams with vector and axial-vector internal propagators. The second term (the derivative of $\mathcal{J}^{\mu}_L$ with respect to $r^\mu$) generates an additional $O(p^0)$ contribution through Goldstone loops.
The combined result can be written in the form:
\begin{eqnarray}\label{eq:Q1result}
\lefteqn{\sum_{ijkl}\mathcal{Q}_L^{ij} \mathcal{Q}_R^{kl}\;
[Q_{1}^{LR}]_{ijkl} =
\frac{3}{32\pi^2}\, \langle \mathcal{Q}_{L} U \mathcal{Q}_{R} U^\dagger\rangle } &&\!\!
\nonumber\\
&&\times 
\int_0^\infty dQ^2\,\left\{ \sum_V\frac{F_{V_i}^2 M_{V_i}^4}{M_{V_i}^2+Q^2} - \sum_A\frac{F_{A_i}^2 M_{A_i}^4}{M_{A_i}^2+Q^2}
\right\} ,\quad
\end{eqnarray}
where the Weinberg conditions \cite{Weinberg:1967kj}
\begin{eqnarray}\label{eq:Weinberg}
\sum_i \left( F_{V_i}^2-F_{A_i}^2\right) & =& F^2\, ,
\nonumber\\
\sum_i \left( M_{V_i}^2 F_{V_i}^2-M_{A_i}^2 F_{A_i}^2\right) & =& 0\, ,
\end{eqnarray}
have been used in order to simplify the final expression. Eq.~\ref{eq:Q1result} agrees with the result obtained in \cite{Knecht:1998sp}, using the alternative Proca description of spin-1 fields.
Performing the integration in the low-energy region $0\le Q^2\le \mu^2$ one obtains the result for $\left.\Delta C_Z\right|_{Q_1^{LR}}$ in Eq.~\ref{q1}.

\bibliographystyle{utphys}
\bibliography{main}

\end{document}